# Towards a Critical Open-Source Software Database

Tobias Dam
tobias.dam@fhstp.ac.at
St. Pölten University of Applied
Sciences
Austria

Lukas Daniel Klausner
mail@l17r.eu
St. Pölten University of Applied
Sciences
Austria

Sebastian Neumaier
sebastian.neumaier@fhstp.ac.at
St. Pölten University of Applied
Sciences
Austria

## ABSTRACT

Open-source software (OSS) plays a vital role in the modern software ecosystem. However, the maintenance and sustainability of OSS projects can be challenging. In this paper, we present the CrOSSD project, which aims to build a database of OSS projects and measure their current project "health" status. In the project, we will use both quantitative and qualitative metrics to evaluate the health of OSS projects. The quantitative metrics will be gathered through automated crawling of meta information such as the number of contributors, commits and lines of code. Qualitative metrics will be gathered for selected "critical" projects through manual analysis and automated tools, including aspects such as sustainability, funding, community engagement and adherence to security policies. The results of the analysis will be presented on a user-friendly web platform, which will allow users to view the health of individual OSS projects as well as the overall health of the OSS ecosystem. With this approach, the CrOSSD project provides a comprehensive and up-to-date view of the health of OSS projects, making it easier for developers, maintainers and other stakeholders to understand the health of OSS projects and make informed decisions about their use and maintenance.

## CCS CONCEPTS

• **Software and its engineering** → **Software libraries and repositories**; **Open source model**; • **Security and privacy** → Software and application security.

## KEYWORDS

open-source software, open-source health, software security, quality monitoring

**ACM Reference Format:**
Tobias Dam, Lukas Daniel Klausner, and Sebastian Neumaier. 2023. Towards a Critical Open-Source Software Database. In *Companion Proceedings of the ACM Web Conference 2023 (WWW '23 Companion), April 30–May 4, 2023, Austin, TX, USA.* ACM, New York, NY, USA, 4 pages. https://doi.org/10.1145/3543873.3587336

## 1 INTRODUCTION

Open-source software (OSS) has become a vital aspect of modern technology, powering a wide range of software applications and providing the infrastructure for many services on the internet. However, sustainability and maintenance of OSS projects can be a significant challenge. To address this, the Critical Open-Source Software Database (CrOSSD) project aims to identify and evaluate the "health" of critical OSS projects, providing a comprehensive overview by using various suitable metrics and (partially automated) analyses. Suitable metrics will include features such as stability, resilience, security and compliance, and will include both quantitative and qualitative metrics (see Section 6 for details).

The result will be a platform monitoring and evaluating a large corpus of OSS projects, which will allow different kinds of stakeholders (the project owners themselves, the OSS community as a whole, other developers, companies, funding authorities, government agencies, non-profit foundations, NGOs, …) to make informed decisions both regarding which projects to support (with funding, labour, cooperation, …) to ensure stability and resilience of critical OSS projects and regarding which projects to use in their own code. This paper presents the methodology and the first steps of CrOSSD, including community engagement processes, database development, metrics and analysis implementation as well as the implications of the project for the OSS community.

There are some extant projects towards that goal (metrics, best practices, some even with scores similar to our basic idea; see Section 2 for details), but none with as comprehensive an understanding of project health as ours (including stability, resilience, security, compliance and more); moreover, none of the existing projects offer continuous monitoring and status reports. We want to offer our current research design for CrOSSD up for discussion at this early stage to garner feedback from the OSS community, developers and other stakeholders and to open ourselves up to critical review to ensure that our approach is the best possible fit for the community's needs and requirements.

We make the following contributions:

- We give a comprehensive overview over CrOSSD's research design, methodology and objectives.
- We present existing projects with a similar goal and explain how we will build upon, complement and surpass them.
- We give an outlook on CrOSSD's planned development and different project phases and describe the envisioned architecture of the platform.
- We explain our structured process for eliciting and incorporating feedback and critique from the OSS community.
- We present a high-level collection and categorisation (i. e. a preliminary taxonomy) of proposed metrics to be used in CrOSSD.

The paper is structured as follows: We first give a comprehensive overview over related work in Section 2, then explain our methodology and research design for CrOSSD in Section 3; the community engagement processes in Section 4; the project architecture and data sources in Section 5; and the types of metrics we will use in Section 6. We conclude the paper with an outlook on future work in Section 7.

## 2 RELATED WORK

Various approaches have been proposed to assess the health and sustainability of OSS projects, including the use of metrics and (automated) analysis of project activity.

One approach is the use of quantitative metrics such as the number of contributors, commits or lines of code to measure the health of an OSS project. For example, the OpenHub project [2] provides a set of metrics and a web interface to visualise them, allowing users to analyse the activity and health of OSS projects written in Python. Similarly, the Black Duck Open Hub[1] (formerly known as Ohloh) provides detailed information on the code and community of OSS projects, including metrics such as the number of contributors, commits and lines of code.

The Open Source Security Foundation (OpenSSF) oversees community activities, working groups and training as well as best practices for OSS projects.[2] The OpenSSF has launched several initiatives to increase the security of OSS projects, such as the criticality score and the Security Metrics Project. The OpenSSF's criticality score[3] defines the influence and importance of a project; it takes values from 0 (least critical) to 1 (most critical) and takes into account parameters such as commits and updates. The Security Metrics Project[4] aims to collect and provide security metrics about open-source projects. It focuses on quantitative metrics, whereas our approach will also include qualitative metrics.

Another example is the Community Health Analytics Open Source Software (CHAOSS)[5] project. CHAOSS aims to provide metrics and analytics to evaluate the health of OSS communities and projects. Besides more generic metrics, CHAOSS also takes inclusion, diversity and risks of OSS projects into account, and in general has a stronger focus on community aspects of project health. Goggins et al. [3] conducted a four-year research study and provide insight into the work of the CHAOSS project, including both the metrics used as well as the open-source implementation Augur.[6] Since they elaborate on the OSS community's as well as other stakeholders' opinions (collected during their field study) regarding suitable definitions of health and sustainability, their work might prove useful for selecting and adapting our own metrics.

CHAOSS and OpenSFF are both complementary to our project, as they mainly define metrics (many of which are not easily automatable as defined) and no explicit OSS database exists based on either project. We will carefully evaluate to what degree we can build upon and reuse concepts and definitions from these projects.

Various research articles [1, 6] focus on the collection and provision of security-related metrics and information on open-source projects. Although the health of OSS projects comprises several different aspects, security is a vital part, affecting both integrity and sustainability. Therefore, we will use their findings and methodology as groundwork for our own research.

In contrast to existing work, CrOSSD will provide a more holistic view of the state of OSS projects by including a variety of metrics, both quantitative and qualitative, and by conducting (automated) analysis of metadata, dependencies, sustainability and funding, activity indicators, security policies, etc. to evaluate the overall health of OSS projects. Furthermore, CrOSSD will set up a monitoring and evaluation platform for a large corpus of OSS projects, thus providing a practically usable source of information for a number of different stakeholder types to serve as a base for informed decision-making and allowing for a more comprehensive and up-to-date view of the health of the OSS ecosystem as a whole.

## 3 METHODOLOGY

In the sequel, we outline our methods and first approaches, including our steps to engage the OSS community, our envisioned platform architecture and a preliminary taxonomy of health metrics.

Quantitative metrics will be gathered through automated crawling of meta information such as the number of contributors, commits and lines of code. This information can be collected from publicly available sources such as GitHub (or GHTorrent), GitLab and other code-hosting platforms, language-specific repositories such as PyPI or aggregators such as Libraries.io. Additionally, we also consider the dependencies of OSS projects to identify any potential issues with compatibility or security vulnerabilities.

Qualitative metrics will be gathered through a combination of manual analysis and automated tools; they will be collected for critical projects only – to limit the potentially significant effort of manual assessment – and will include aspects such as sustainability, funding, community engagement and adherence to security policies. Initially, we consider an OSS project to be *critical* if its failure, compromise or disappearance would pose a substantial risk to security, reliability or availability due to its widespread distribution, popularity or number of dependencies; we plan to adapt and fine-tune our criticality metric through community engagement, by including the OpenSSF's criticality score, etc.

For automatically collectable data, we plan continuous monitoring and regular updates as well as providing the history of the gathered data. For qualitative data where automation is less feasible, we plan to employ different mechanisms for data acquisition (e. g. surveying the projects, allowing for sourced "push" updates by external contributors, ...) which will be available as explicitly timestamped information.

## 4 COMMUNITY ENGAGEMENT

We will ensure that CrOSSD will be useful to both the OSS community and external stakeholders by setting up structured, continuous engagement, evaluation and input processes to elicit and incorporate feedback and critique.

- We will conduct *qualitative interviews* with OSS community representatives as well as individual and institutional stakeholders (such as industry, funding authorities and government agencies).

---

[1] https://www.openhub.net/
[2] https://bestpractices.coreinfrastructure.org/en/criteria
[3] https://github.com/ossf/criticality_score
[4] https://metrics.openssf.org/ and https://github.com/ossf/Project-Security-Metrics
[5] https://chaoss.community
[6] http://www.github.com/CHAOSS/augur

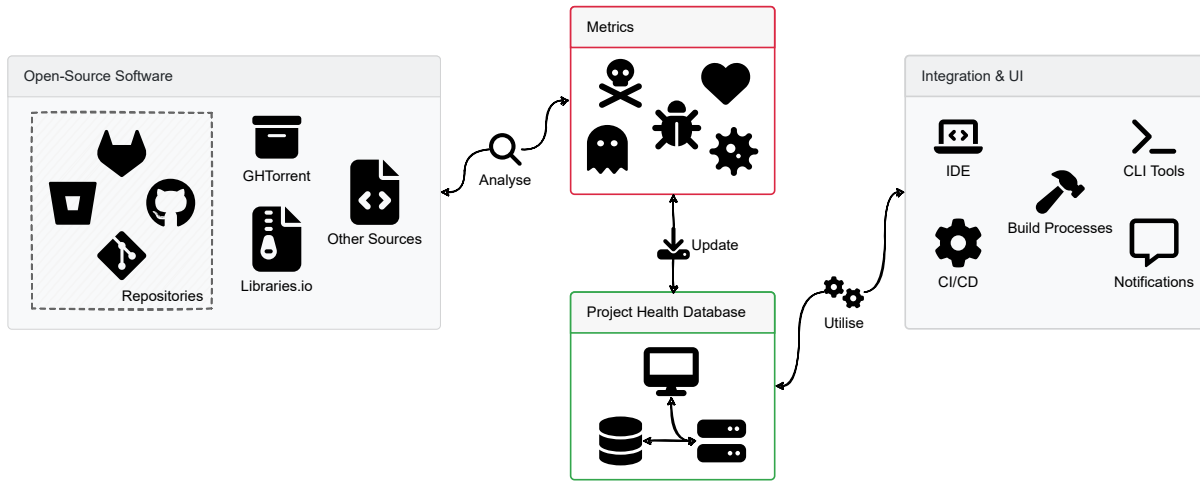

Figure 1: Conceptual architecture of the CrOSSD components.

- We will set up *quantitative surveys* to get a representative picture of requirements, needs and expectations in the OSS community.
- We will contact and work with *open-source community platforms* (such as Github or Gitlab) and existing *collaboration partners* (e. g. Red Hat) to obtain input from their userbases.
- We will use both existing relevant *events, conferences and workshops* (e. g. the Open Source Summit Europe) as well as meetings that we will organise ourselves to network and hold in-depth discussions and focus groups.

We consider these community engagement activities of fundamental importance to ensure the acceptance, sustainability and long-term relevance of CrOSSD.

## 5 ARCHITECTURE

The architecture displayed in Figure 1 consists of several components that work together to collect, analyse and present the data:

*Open-Source Software:* The data collection component is responsible for gathering various types of meta information about OSS projects. Potential data sources include:

- *Code-hosting platforms:* The project will gather data from publicly available *repositories* such as GitHub, GitLab and other code-hosting platforms. These data include meta information about the OSS projects, such as the number of contributors, commits and lines of code.
- *Existing collections and crawls:* Some existing datasets such as GHTorrent [4] offer a collection of metadata from the GitHub platform. GHTorrent provides all the data available on GitHub, including issues, pull requests and commits. A similar data source is Libraries.io, which crawls and monitors a range of open-source package managers and provides APIs to retrieve the collected data.
- *Financial reports and grants:* Data on funding for OSS projects through publicly available financial reports and grants can be used to evaluate the sustainability of projects.
- *Community engagement data:* Data on community engagement, such as mailing list activity and number of pull requests, can be used to evaluate the level of activity within the project and the level of engagement of a project's users.
- *Security vulnerabilities databases:* We will consider data on known security vulnerabilities within the OSS projects from publicly available databases such as the Common Vulnerabilities and Exposures (CVE) database [5].

*Metrics:* The metrics assessment component is responsible for analysing the collected data. This component generates scores reflecting the health of the OSS projects in a number of different domains; scoring and analysis processes will be developed in collaboration with experts from the OSS community and will be regularly reviewed and updated to ensure they remain relevant and accurate.

*Project Health Database:* The project health database component is responsible for continuously monitoring and updating the metrics' assessment.

*Integration and UI:.* The goal of CrOSSD is to provide several ways of accessing and integrating the results. First, the results of the analysis will be displayed on a user-friendly *web platform* which allows users to view the health of individual OSS projects as well as the overall health of the OSS ecosystem. Users can filter and search for projects based on different criteria, such as programming language or type of licence, and also view detailed information about the project, including metrics, scores and analyses.

Further, we aim to provide *APIs and push-based services* to allow the integration of the CrOSSD platform in IDEs, CI/CD and build processes. Push-based alerts or notifications can be sent to interested users in case critical situations are detected in any of the OSS projects they have watchlisted.

The platform is designed to be scalable and flexible, allowing for the addition of new data sources and analysis techniques as needed. Additionally, the platform is built using open-source technologies to allow for easy integration with other tools and platforms used by the OSS community.

## 6 TAXONOMY OF METRICS

The metrics used in CrOSSD can be distinguished along two main dimensions: quantitative vs. qualitative and by focus (i. e. security, activity and relevance metrics).

Quantitative metrics are numerical measurements of various aspects of the OSS project, e. g.:

- *Number of contributors:* A high number of people contributing to an OSS project and frequently pushing commits can indicate a healthy and active community.
- *Number of commits:* A high number of code changes can indicate an active development process, although less activity can also merely indicate an established, older project with a relatively stable codebase.
- *Dependency analysis:* Measuring the number and type of dependencies the project has can help identify potential issues with compatibility or security vulnerabilities.

Qualitative metrics, on the other hand, are those which may be based on subjective evaluations of the OSS project, rely on manual data collection or (at least partial) self-reporting or require manual analysis of data to derive them, e. g.:

- *Compliance:* This includes adherence to legal and regulatory requirements, such as using appropriate licences, supplying appropriate metadata or respecting data protection law.
- *Funding:* This metric measures the support the project has from grant agencies or institutional supporters (e. g. by supplying labour from employees).
- *Sustainability*: This refers to a project's ability to be maintained and developed over time, including aspects such as governance and community engagement.

Security metrics are designed to evaluate the security of the OSS project, such as:

- *Security policies:* This reflects adherence to security best practices within the project, such as code reviews and testing.
- *Vulnerabilities:* This metric measures the number and severity of known security vulnerabilities within the project and how long newly found vulnerabilities remain unfixed.

Activity metrics are meant to reflect the level of activity within the OSS project, such as:

- *Activity indicators:* This metric measures the level of activity within the project, including the number of pull requests and mailing list activity.
- *User engagement:* This metric measures the level of engagement of the users with the project, including the amount of feedback provided by the users as well as the number of forks.

Relevance metrics are designed to capture an OSS project's relevance, these include:

- *Popularity:* This metric measures the popularity of the project, including the number of downloads, installs and references.
- *Industry adoption:* This metric measures the number of companies or organisations that are using the project.
- *Contribution diversity*: This metric measures how many different stakeholders or interest groups are actively involved in maintaining the project.

The specific metrics used will be defined in collaboration with experts from the OSS community and will be regularly reviewed and updated to ensure they remain relevant and accurate.

## 7 CONCLUSION

In conclusion, CrOSSD aims to build a comprehensive database of critical open-source software (OSS) projects and measure their current "health" status through (partially automated) analyses. The project uses both quantitative and qualitative metrics, including the number of contributors, commits, lines of code, sustainability, community engagement and adherence to security policies. The data for these metrics are collected from various publicly available sources such as code-hosting platforms, financial reports, grants, community engagement data and security vulnerabilities databases.

For next steps, we will focus on the development and implementation of the platform as well as gathering input from the OSS and academic community to ensure that the metrics used are relevant and accurate. A particular goal at this state is establishing collaborations with different partners from academia, industry and the public sector to increase the impact of our work.

To foster interoperability of our approach, we plan to develop an ontology of metrics to be published as FAIR data. We also aim to expand the platform to cover OSS projects from various countries and languages to make it more inclusive and representative of the global OSS ecosystem.

## ACKNOWLEDGMENTS

This research was funded by the netidee Call 17 project 6252 "Towards a Critical Open-Source Software Database". The financial support by the Internet Foundation Austria is gratefully acknowledged. We are grateful to Philipp Haindl and Torsten Priebe for suggesting numerous improvements to both the content and the presentation of this paper.